\documentclass[prl,aps,twocolumn,superscriptaddress,amsmath,amssymb]{revtex4}

\usepackage{graphicx}
\usepackage{dcolumn}
\usepackage{bm}
\usepackage{color}
\usepackage{pdfsync}
\usepackage[a4paper,hmargin={1.5cm,1.5cm},vmargin={1.5cm,2.5cm}]{geometry} 
\usepackage{hyperref}

\begin{document}


\title{Strong Coupling of a Quantum Oscillator to a Flux Qubit at its Symmetry Point }

\author{A.~Fedorov}
\altaffiliation{Present address: Department of Physics, ETH Zurich, CH-8093, Zurich, Switzerland}
\email{fedoroar@phys.ethz.ch}
\affiliation{Kavli Institute of Nanoscience, Delft University of Technology, PO Box 5046, 2600 GA Delft, The Netherlands}

\author{A.~K.~Feofanov}
\affiliation{Physikalisches Institut and DFG Center for Functional Nanostructures (CFN) Karlsruhe Institute of Technology, Wolfgang-Gaede-Str. 1, D-76131 Karlsruhe, Germany}
\affiliation{Kavli Institute of Nanoscience, Delft University of Technology, PO Box 5046, 2600 GA Delft, The Netherlands}

\author{P.~Macha}
\affiliation{Institute of Photonic Technology, P.O. Box 100239, D-07702 Jena, Germany}
\affiliation{Kavli Institute of Nanoscience, Delft University of Technology, PO Box 5046, 2600 GA Delft, The Netherlands}

\author{P. Forn-D\'iaz}
\affiliation{Kavli Institute of Nanoscience, Delft University of Technology, PO Box 5046, 2600 GA Delft, The Netherlands}

\author{C. J. P. M.~Harmans}
\affiliation{Kavli Institute of Nanoscience, Delft University of Technology, PO Box 5046, 2600 GA Delft, The Netherlands}

\author{J. E.~Mooij}
\affiliation{Kavli Institute of Nanoscience, Delft University of Technology, PO Box 5046, 2600 GA Delft, The Netherlands}

\begin{abstract}
A flux qubit biased at its symmetry point shows a minimum in the energy splitting (the gap), providing protection against flux noise. We have fabricated a qubit of which the gap can be tuned fast and have coupled this qubit strongly to an LC oscillator. We show full spectroscopy of the qubit-oscillator system and generate vacuum Rabi oscillations. When the gap is made equal to the oscillator frequency $\nu_{osc}$ we find the largest vacuum Rabi splitting of $\sim0.1\nu_{\rm osc}$. Here being at resonance coincides with the optimal coherence of the symmetry point.
\end{abstract}


\maketitle

Superconducting qubits coupled to quantum oscillators have demonstrated a remarkable richness of physical phenomena in the last few years. After the first reports of coherent state transfer and strong coupling~\cite{Delft side-bands,wallraff}, we have witnessed a rapid development of the field called circuit quantum electrodynamics (CQED) using high quality superconducting oscillators  in realizing quantum gates~\cite{majer}, algorithms~\cite{dicarlo} as well as non-classical states of light and matter in artificially fabricated structures~\cite{wallraff2,Martinis2}. Among the different implementations the transmon~\cite{wallraff,majer,dicarlo,wallraff2} and the phase qubit~\cite{Martinis2} dominated this development. With flux qubits the avoided crossing between qubit and oscillator level was observed~\cite{jena,Johansson} and the coherent single-photon exchange between qubit and oscillator was demonstrated~\cite{Johansson}.
However the, coherence of the flux qubit is optimally preserved only in the symmetry point for flux bias, where the energy splitting is minimal. This minimal splitting  ($h\Delta$) is called the gap and depends (exponentially) on the properties of the Josephson junctions. Therefore, the gap is hard to control in fabrication and it is impossible to make it coincide with a fixed oscillator frequency.  We now have developed a flux qubit of which the gap $\Delta$ can be tuned over a broad range on sub-ns time scales~\cite{floor}.
With the use of this control we demonstrate strong coupling of a flux qubit with good coherence to a lumped-element LC oscillator, showing fast and long-lived vacuum Rabi oscillations.


Parameters of the superconducting qubits can be to a large extent chosen in the design phase. For strong coupling, where the interaction strength $g$ exceeds the cavity and qubit loss rates, the rotating-wave approximation (RWA) can be applied and the system can be described by a Jaynes-Cummings type Hamiltonian. If $g$ approaches the qubit or oscillator frequencies the RWA no longer holds, leading into the ultra-strong coupling regime~\cite{Bourassa, Pol}.
For a flux qubit the ratio $g/\nu_{\rm osc}$ can be an order of magnitude larger than for charge and phase qubits~\cite{coupling limit}, while these latter devices have a coupling that can be several orders of magnitude larger than the atom-light interaction energy~\cite{wallraff}. For good coherence, operating the qubit at its spectral symmetry point is required. Therefore, experimentally combining galvanic coupling of oscillator and flux qubit with this symmetry point operation provides a major step forward in the development of CQED systems. For the flux qubit at the symmetry point the anharmonicity (distance between 2nd and 3rd level relative to qubit splitting) is very high, allowing very fast operation without quantum leakage.

\begin{figure}
\includegraphics[width=90mm]{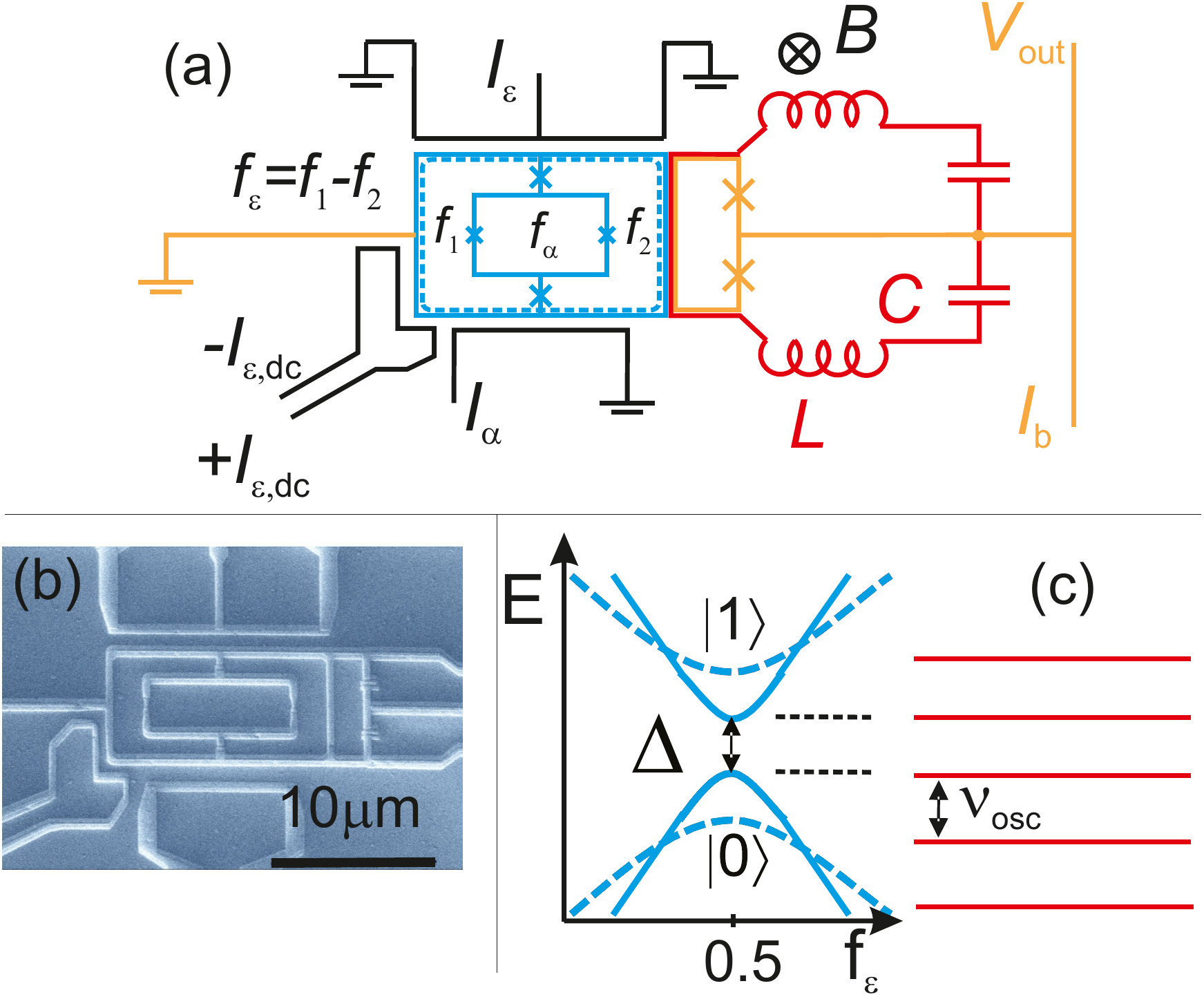}
\caption{\label{fig:scheme}(color online). (a) Circuit schematics: the tunable gap flux qubit (green) coupled to a lumped element superconducting LC oscillator (red) and controlled by the bias lines $I_\epsilon$, $I_{\epsilon,dc}$, $I_\alpha$ (black). The SQUID (blue) measures the state of the qubit. The gradiometer loop (emphasized by a dashed line) is used to trap fluxoids. (b) Scanning Electron Micrograph (SEM) of the sample.
(c) Energy diagram of the qubit-oscillator system. The minimum of energy splitting of the qubit $\Delta$ is reached at the symmetry point when one fluxoid is trapped in the gradiometer loop and the difference in magnetic fluxes $f_\epsilon\Phi_0$ is $0$ controlled by $I_\epsilon$ and $I_{\epsilon,dc}$. By controlling the flux $f_\alpha\Phi_0$ with $I_{\alpha}$ and uniform field $B$ one can tune $\Delta$ in resonance with oscillator frequency $\nu_{\rm osc}$.
 }
\end{figure}

The investigated system is represented in Fig.~1. The flux qubit has a gradiometric topology~\cite{floor, Koch} by having the Josephson junctions that form the qubit symmetrically attached to the circumference loop as shown in Fig.~1(a,b); this loop is also employed to trap fluxoids (or 2$\pi$-phase-winding numbers)~\cite{fluxon trap}. To obtain a tunable-gap qubit the two center junctions form a SQUID structure, where the flux $f_\alpha\Phi_0$ {\it in situ} sets the effective critical current and in this way the qubit gap $\Delta$~\cite{floor}. The gap covers nearly two decades from 150 MHz to 12 GHz, providing full frequency control relative to the oscillator at 2.723~GHz (see Fig.~1(c)). The Hamiltonian of the flux qubit can be written as
${H_{\rm qb}}=-
h(\epsilon \sigma_z+\Delta \sigma_x)/2$,
where $\sigma_z$ and $\sigma_x$ are Pauli matrices written in the persistent current states basis; $h\epsilon$ is the magnetic energy bias  $h\epsilon(f_\epsilon,f_\alpha)=2I_p(f_\alpha)f_\epsilon\Phi_0$, with $I_p$ being the circulating current in the qubit and $2f_\epsilon\Phi_0=(f_1-f_2)\Phi_0$ describing the difference in flux in the two loop halves of the gradiometer (Fig.~1(a)). Qubit excitation is obtained by the quadrupolar magnetic field generated by current in the symmetrically-split $I_\epsilon$  line, acting on the qubit flux $f_\epsilon\Phi_0$. Similarly, the line $I_\alpha$  together with the homogeneous field $B$ generated by an
external coil, modulates $f_\alpha\Phi_0$  and changes $\Delta$. The structural symmetry suppresses crosstalk, implying a fully selective control.

\begin{figure}
\includegraphics[width=90 mm]{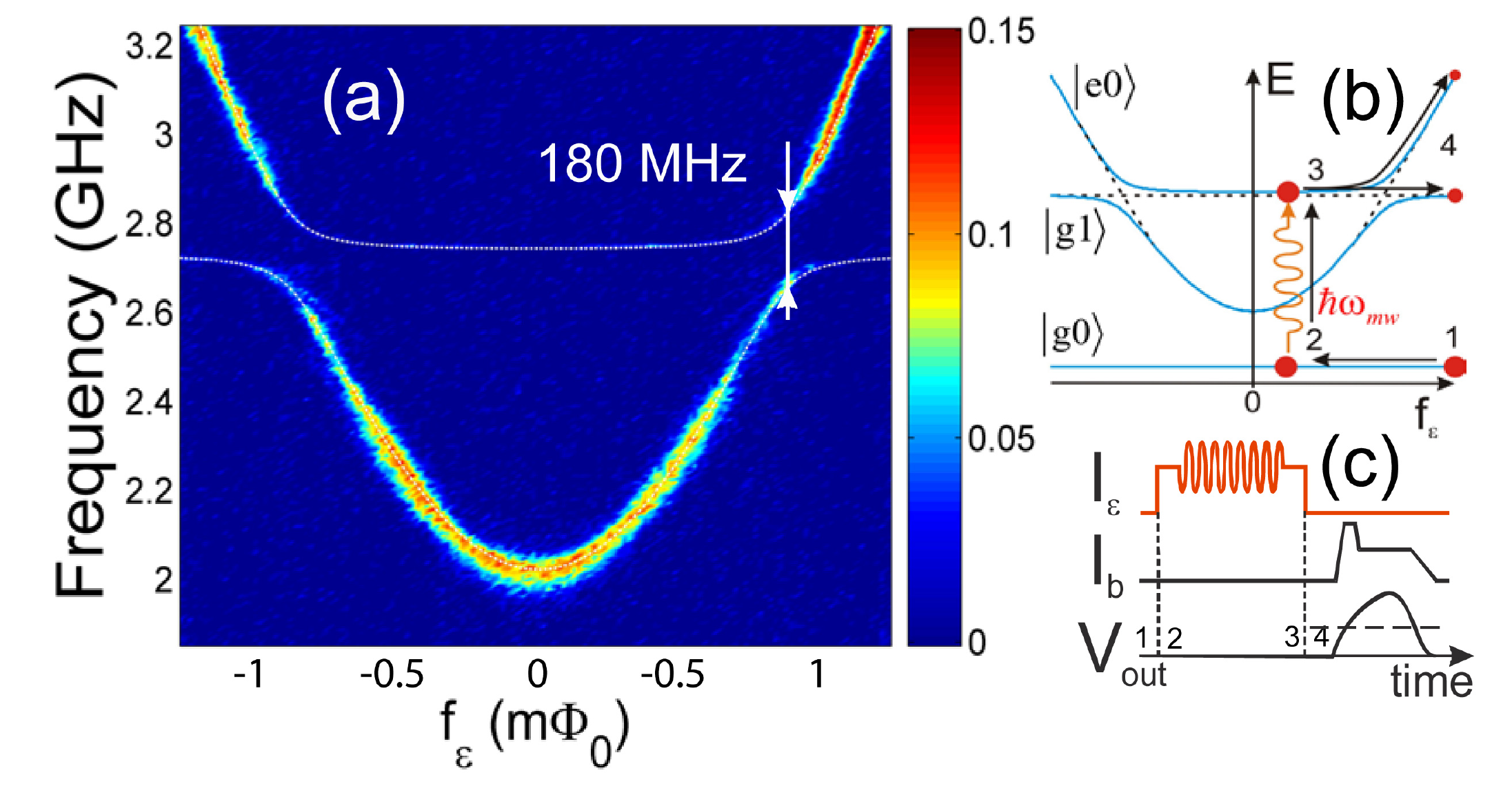}
\caption{\label{fig:spectrum1}(color online).
 (a) Schematic representation of the control and measurement pulses to perform spectroscopy. (b) Diagram of Landau-Zener transitions  transferring the excitation of the oscillator to the qubit. (c) MW frequency vs  $f_\epsilon$ (controlled by the amplitude of the current pulse $I_\epsilon$). The color indicates the switching probability of the SQUID minus 0.5. The white dotted line is obtained from Eq.~(\ref{total H1}) with $\Delta=2.04$~GHz, $I_p=420$~nA. The vacuum Rabi splitting of 180 MHz corresponds to the effective qubit-oscillator coupling strength reduced by $\sin\eta$.}
\end{figure}

The qubit states are detected with a dc-SQUID which is coupled to the qubit with a shared part of a wire of length $l=6~\mu$m, width $w=350$~nm, and thickness $t=70$~nm leading to a mutual qubit-SQUID inductance $M\simeq5.5$~pH. Half of $M$ is provided by a kinetic inductance of the shared part $L_K\sim l/(tw)$ which can be easily made even larger than the geometric contribution~\cite{Pol}.
At zero bias current the noise induced by the readout circuitry is decoupled from $f_\alpha$ and $f_\epsilon$ due to the SQUID-qubit geometry. The junctions of the SQUID are shunted with two on-chip parallel plate capacitors of $C=8$~pF reducing the plasma frequency to $\nu_{\rm p}\approx1.3$~GHz.


The inductances of wires $L$ and capacitors $C$ in series form a lumped element LC oscillator with $\nu_{\rm osc}=1/\left(2\pi\sqrt{2L(C/2)}\right)=2.723$~GHz.
 The oscillator is coupled to the qubit via the same shared part of the wire with the qubit-SQUID mutual inductance $M\simeq5.5$~pH (since $M$ is much smaller than the Josephson inductances of the SQUID junctions most of the current flows through the shared part). Unlike the plasma resonance of the SQUID the connections of the  oscillator to the external circuit occurs in
the voltage nodes of the resonance mode. Thus the oscillator quality
factor $Q$ is not severely affected by the external impedance,
reaching $Q\sim6000$ for strong excitation and a few hundreds at low
photon number. Being designed as part of the readout circuit, it was not
optimized for high $Q$ performance.

All structures excluding the bottom plate of the capacitors $C$ were fabricated in the same layer of aluminum using standard lithography techniques~\cite{Patrice}. The bottom plates of the capacitors were fabricated in a separate layer followed by a plasma oxidation step resulting in a thin layer of Al-AlO$_x$-Al used as the dielectric of the capacitor. The experiment was conducted in a dilution refrigerator at its base temperature of $20$~mK.

The interaction between the qubit and the oscillator can be described by $H_{\rm int}=hg(a+a^{\dagger})\sigma_z$  written in the basis of the persistent current states,  where $g=M I_p I_0$ is the coupling strength, $I_0= \sqrt{h \nu_{\rm osc}/(4L)}$ is the measure for zero-point current fluctuations,  $a^{\dagger}$, $a$  are photon creation and annihilation operators of the oscillator defined in the oscillator Fock space ${|n\rangle}$. In the energy eigenstates of the qubit, $\{|g\rangle,\,|e\rangle\}$,  the system Hamiltonian reads
\begin{eqnarray}
H=\frac{h\nu_{\rm qb}}{2}\sigma_z&+&h\nu_{\rm osc} \left(a^{\dagger}a+\frac{1}{2}\right)\nonumber\\
&+&h g\left(\cos\eta \, \sigma_z-\sin\eta \, \sigma_x\right) (a+ a^{\dagger}),
\label{total H1}
\end{eqnarray}
where $h\nu_{\rm qb}\equiv h\sqrt{\Delta^2+\epsilon^2}$ is the qubit energy splitting and $\tan \eta\equiv\Delta/\epsilon$. In the following we examine two representative cases $\Delta= \nu_{\rm osc}$ and  $\Delta< \nu_{\rm osc}$. In particular, in the former case we concurrently establish maximum coupling and maximum coherence, favored by the coincidence of the resonance condition with the qubit symmetry point.

The spectroscopy of the system was performed with the protocol sketched in
Fig.~2(a,b). First we set the gap of the qubit with the external
magnetic field $B$ and applied a dc offset to $I_{dc, \epsilon}$ to tune
the qubit frequency to $\nu_{\rm qb}=9$~GHz. In
the second step we applied a square current pulse in $I_{\epsilon}$, tuning $\nu_{\rm qb}$ to the required frequency, combined with a
microwave (MW) excitation.  After each
excitation pulse the qubit was returned to $\nu_{\rm qb}=9$~GHz and a
short bias current pulse $I_b$ was applied to the SQUID for
measurement of the qubit state~\cite{Patrice}.

By measuring the qubit away from its symmetry point we benefit from the high expectation value of the circulating  current of the qubit eigenstates and a long relaxation time $T_1$, gradually increasing from $T_1\simeq1.5$~$\mu$s in the symmetry point to $T_1>4$~$\mu$s at $\nu_{\rm qb}\simeq9$~GHz with $\Delta\simeq 2$~GHz.

 Figure~2(c) shows the spectrum of the system for $\Delta<\nu_{\rm osc}$. In order to be resonant with the oscillator the qubit has to be tuned away from its symmetry point. The clear observation of the level anti-crossing with the largest level splitting of $180$~MHz (vacuum Rabi splitting) confirms that the system is in the strong cQED regime. To observe the spectral line of the oscillator we use a Landau-Zener transition at the anti-crossing of the qubit and oscillator energies. A passage through the anti-crossing region performed with a dc-shift pulse on $I_\epsilon$ with a rise time of 4~ns is found to lead to $\sim25\%$ probability of an oscillator photon to be converted to the excited state of the qubit. The latter can be detected by the SQUID and the oscillator line becomes visible on spectrum.


\begin{figure}
\includegraphics[width=87mm]{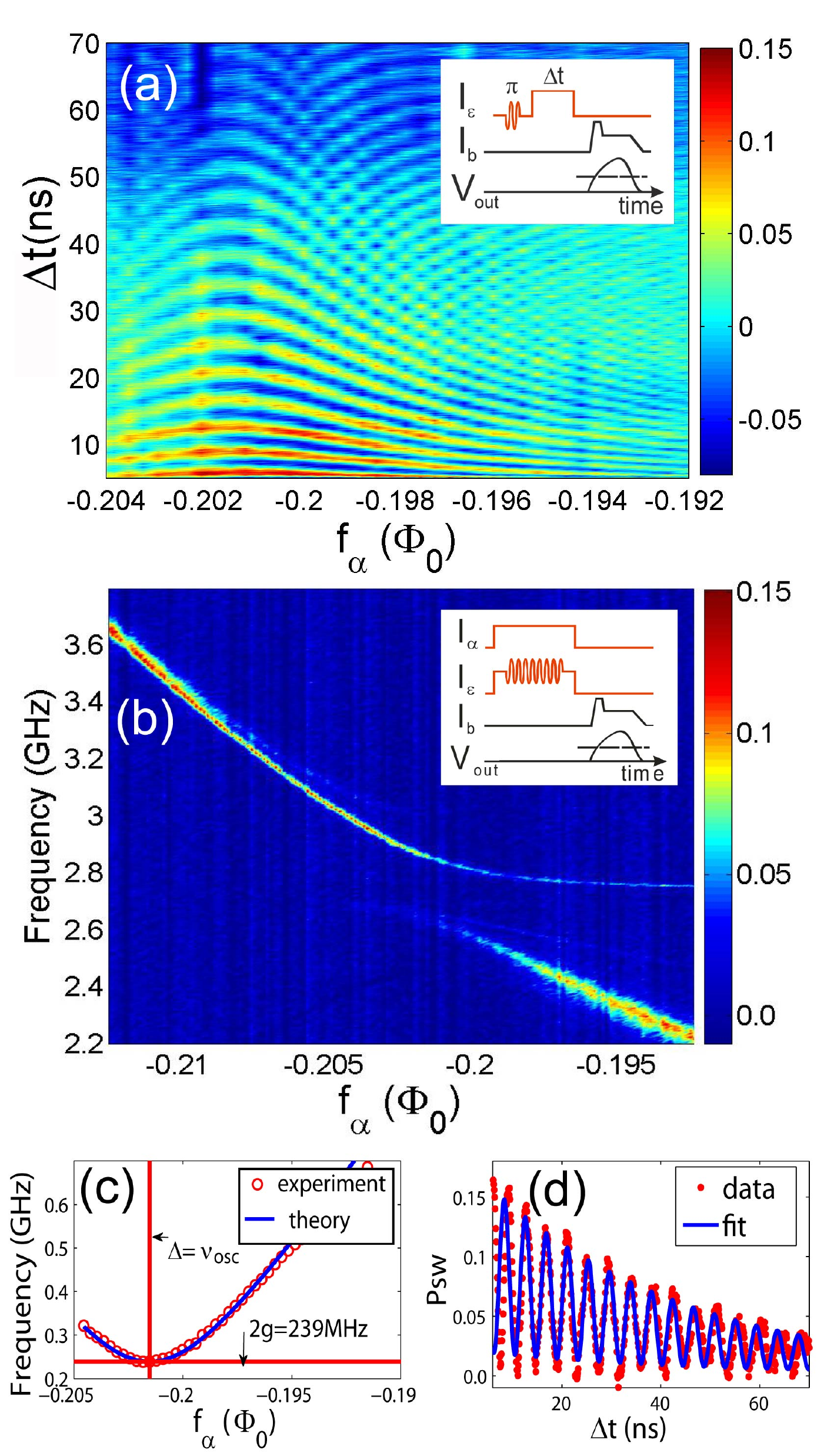}
\caption{\label{fig:rabi_sym}(color online).
Vacuum Rabi oscillations (a) and MW frequency (b) vs magnetic $f_\alpha$. In the experiment the qubit was kept in its symmetry point ($\epsilon=0$) by appropriately adjusting the amplitude of the current pulse $I_\epsilon$ while $\Delta$ was changed by $f_\alpha$ with use of external magnetic field $B$ (a) or by applying the current pulse $I_\alpha$ for fixed $B$ (b). The color scale shows the switching probability of the SQUID minus 0.5.
(c) Frequency of the vacuum Rabi oscillations extracted from data (a) and theoretical estimation (blue line) from Eq.~(\ref{Rabi frequency}) as a function of $f_\alpha$. The minimum in $\nu_R$ determines the bare qubit-oscillator coupling $2g$ and corresponds to the resonance conditions $\Delta=\nu_{\rm osc}$. (d) Single trace of the vacuum Rabi oscillations for $\Delta\simeq \nu_{\rm osc}$.
}
\end{figure}

The sequence of operations to observe the vacuum Rabi oscillations starts by tuning the qubit gap $\Delta$  into the vicinity of $\nu_{\rm osc}$, setting $f_\alpha$  by the external magnetic field.  The qubit is tuned to  $\nu_{\rm qb}=7$~GHz by $f_\epsilon$, and a $\pi$-pulse is applied to excite the qubit. Subsequently, the qubit is taken to the symmetry point by means of a fast 0.3~ns rise time pulse. As the qubit energy changes fast relative to the coupling strength $g$, this transfer is non-adiabatic. The qubit is kept here for a time $\Delta t$, then returned fast to the 7~GHz level and finally read out. While the qubit is in the symmetry point, qubit and oscillator coherently exchange the excitation with a frequency that is determined by the coupling and the detuning $\delta\nu=\Delta-\nu_{\rm osc}$ according to~\cite{book}
\begin{equation}
\label{Rabi frequency}
\nu_{R}\cong\sqrt{4g^2+\delta\nu^2}.
\end{equation}
The vacuum Rabi oscillations are shown in Fig.~3(a). For each value of $f_\alpha$  (and therefore $\Delta$), the probability to find the qubit in one of its eigenstates oscillates as a function of  $\Delta t$ with a frequency that is minimal for  $f_\alpha\cong-0.202$, the point where $\Delta=\nu_{\rm osc}$. Fig.~3(b) shows the spectrum as a function of  $f_\alpha$, with the avoided crossing clearly visible. From the slope the value $d\Delta/df_\alpha\approx 69.5$~GHz/m$\Phi_0$  can be determined, which is used to estimate $\delta\nu$ as $\delta\nu=(d\Delta/df_\alpha)df_\alpha$. By fitting to Eq.~(\ref{Rabi frequency}) the bare coupling $2g$ is found to be $239$~MHz.

We now focus on the most interesting on-resonance regime with $\Delta=\nu_{\rm osc}$.
From (\ref{total H1}) one can see that here the qubit-oscillator coupling is fully transversal $\eta=\pi/2$, making the effective coupling attain its maximum value $g$. The measurement of the spectrum, shown in Fig.~4(a), indeed exhibits the maximum vacuum Rabi splitting of $239$~MHz corresponding to the highest photon exchange rate between oscillator and qubit.

Interestingly, Landau-Zener transitions now change qualitatively: after the passage through the anti-crossing the energy of the state $|e0\rangle$ remains higher than that for $|g1\rangle$ making the qubit and the oscillator almost fully exchange their populations and creating strong asymmetry in the visibility of the spectral lines in Fig.~4(a) (for $\Delta<\nu_{\rm osc}$, see Fig.~2(b,c) where the qubit and the oscillator tend to retain their populations).

In Fig.~4(b) we demonstrate vacuum Rabi oscillations for different $f_\epsilon$.
Taking into account only $|0,1\rangle$ oscillator states the Rabi frequency can be found analytically from (\ref{total H1}) as
\begin{equation}
\label{Rabi freq general}
\nu_R=\left( 4g^2+\nu_{\rm osc}^2+\nu_{\rm qb}^2-2{\sqrt{4g^2\epsilon^2+\nu_{\rm qb}^2\nu_{\rm osc}^2}} \right)^{1/2},
\end{equation}
which explains the measured data as shown in Fig.~4(c). Note that Eq.~(\ref{Rabi freq general}) reduces to Eq.~(\ref{Rabi frequency}) if $\epsilon=0$.


Implementation of the gap control loop might in principle lead to additional decoherence. However, in practice the effect of flux noise in $\alpha$-loop in our design is estimated to be about two orders of magnitude smaller than that of the $\epsilon$-loop. Measuring the qubit in the symmetry point we found $T_2\simeq300$~ns and $T_1\simeq T_{\rm Rabi}\simeq1.5~\mu$s for $\Delta \sim 1.5-6$~GHz. While $T_1$ and $T_{\rm Rabi}$ are in accordance with design values we observed no dependence of $T_2<2T_1$ on $\Delta$ which rules out flux noise in both $f_{\epsilon}$, $f_{\alpha}$ as a limiting decoherence source in the symmetry point~\cite{decoherence analysis}.

Since the qubit is optimally protected from low-frequency flux noise in the symmetry point the vacuum Rabi oscillations show the longest decay time of $\sim40$~ns. This is limited only by the losses in the oscillator, as measured coherence times of the qubit at the symmetry point are much longer~\cite{book}. Out of the symmetry point we measure the usual rapid degradation of the qubit coherence to $T_2\sim15-20$~ns for $\epsilon\gg\Delta$~\cite{nec2} due to flux noise which precludes generation of long-living vacuum Rabi oscillations. Obviously, by using a fully compatible fabrication technology optimized for high $Q$ oscillators~\cite{wallraff,majer,dicarlo,wallraff2} it is possible to achieve a ratio of $g/{\rm max}(T_2^{-1},T_1^{-1}, 2 \pi \nu_{\rm osc}/Q)>100$ necessary for creating qubit-oscillator entanglement with high fidelity.

\begin{figure}
\includegraphics[width=87mm]{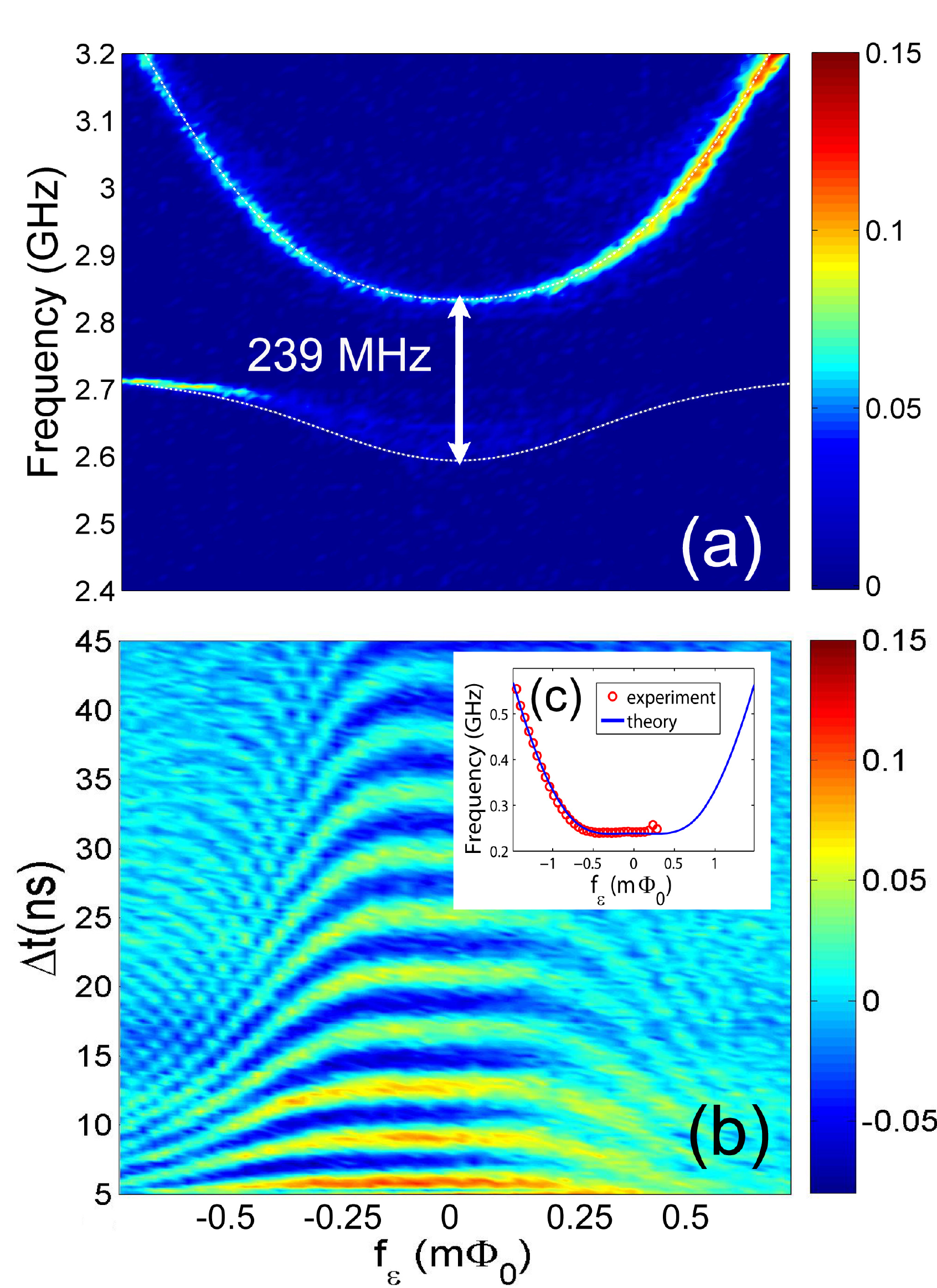}
\caption{\label{fig:rabi_sym_2}(color online).
(a) MW frequency vs $f_\epsilon$. The dotted white line is obtained from Eq.~(\ref{total H1}) with $I_p=400$~nA and $\Delta=\nu_{\rm osc}$. The observed vacuum Rabi splitting is maximal due to fully transverse coupling of the qubit to the oscillator $\eta=\pi/2$.
(b) Vacuum Rabi oscillations  for different values of $f_\epsilon$.
In the experiment $f_\epsilon$ was controlled by the amplitude of the current pulse $I_\epsilon$ while  $\Delta$ was tuned to $\nu_{\rm osc}$ by changing the external magnetic field $B$.
The inset shows $\nu_R$ extracted from data (red circles) and estimated from Eq.~(\ref{Rabi freq general}) (blue line). The color indicates the switching probability of the SQUID minus 0.5.
}
\end{figure}

In summary, we experimentally studied a tunable gap flux qubit coupled galvanically to a superconducting lumped-element LC oscillator.  We measured the avoided level crossings and generated vacuum Rabi oscillations for two representative cases: the gap was tuned substantially below and in resonance with the oscillator frequency.
For the particularly interesting case of $\Delta=\nu_{\rm osc}$ the qubit reaches the resonance conditions in its symmetry point thus combining the two most desired ingredients of the cQED regime: strong coupling and optimal coherence. Here the avoided level crossing attains its maximal value of $ 2g\simeq0.09\cdot\nu_{\rm osc}$ and at the same time the qubit is effectively protected  from 1/f flux noise resulting in the longest and fastest sequence of on-resonant vacuum Rabi oscillations. The interaction strength can be readily  increased reaching the ultra-strong regime $g\sim\{\nu_{\rm osc},\nu_{\rm qb}\}$.

We thank P.~C.~de~Groot and R.~N.~Schouten for useful
discussions. This work was supported by the Dutch NanoNed program, the Dutch Organization for Fundamental Research (FOM), and the EU projects EuroSQIP and CORNER.


\begin{thebibliography}{10}
\bibitem{coupling limit}
Assuming galvanic and capacitive couplings for flux and phase qubits, repspectively, the ratio $g/\nu_{\rm osc}$ for a flux qubit can be $\Phi_0 I_{\rm rms}/(\langle \hat n\rangle 2 eV_{\rm rms})=0.25 R_{K}/Z_0\approx10$ times larger. Here $e$ is a charge of an electron, $R_{K}$ a quantum of resistance, $Z_0$ is the characteristic impedance of the oscillator and $\langle \hat n\rangle$ is a number of charge states.
\bibitem{decoherence analysis}
Detailed analysis of decoherence of the tunable flux qubit will be published elsewhere
\end{thebibliography}
 \end{document}